\documentclass[a4paper,11pt,headings=big]{scrartcl}
\pdfoutput=1
\usepackage{amsmath,amssymb}
\usepackage{feyn}
\usepackage{txfonts}
\usepackage{caption}
\usepackage{subcaption}
\usepackage{color}
\usepackage[colorlinks]{hyperref}
\usepackage{graphicx}
\addtokomafont{disposition}{\rmfamily\boldmath}
\usepackage{enumerate}
\usepackage{slashed}
\usepackage{cite}
\usepackage{multirow}
\usepackage[normalem]{ulem}

\allowdisplaybreaks

\begin{document}

\begin{flushright}
\begin{tabular}{l}
MITP/14-017\\
March 10, 2014
\end{tabular}
\end{flushright}
\vskip1.5cm

\begin{center}
{\LARGE\bf 
Dipole operator constraints on composite Higgs models
}\\[0.8 cm]
{\large%
Matthias K\"onig$^a$,
Matthias Neubert$^{a,b}$,
and David M. Straub$^c$
\\[0.5 cm]
\small
$^a$ \em PRISMA Cluster of Excellence \& Mainz Institute for Theoretical Physics,
Johannes~Gutenberg University, 55099 Mainz, Germany
\\[0.2cm]
$^b$ Department of Physics, LEPP, Cornell University, 
Ithaca, NY 14853, U.S.A.
\\[0.2cm]
$^c$ Excellence Cluster Universe, Technische Universit\"at M\"unchen, Boltzmannstr.~2, 85748~Garching, Germany
}
\\[0.5 cm]
\small
E-Mail:
\tt\href{mailto:m.koenig@uni-mainz.de}{m.koenig@uni-mainz.de},
\tt\href{mailto:neubertm@uni-mainz.de}{neubertm@uni-mainz.de},
\tt\href{mailto:david.straub@tum.de}{david.straub@tum.de}
\end{center}

\bigskip

\begin{abstract}\noindent
Flavour- and CP-violating electromagnetic or chromomagnetic dipole operators in the quark sector are generated in a large class of new physics models and are strongly constrained by measurements of the neutron electric dipole moment and observables sensitive to flavour-changing neutral currents, such as the $B\to X_s\gamma$ branching ratio and $\epsilon'/\epsilon$. After a model-independent discussion of the relevant constraints, we analyze these effects in models with partial compositeness, where the quarks get their masses by mixing with vector-like composite fermions. These scenarios can be seen as the low-energy limit of composite Higgs or warped extra dimensional models. We study different choices for the electroweak representations of the composite fermions motivated by electroweak precision tests as well as different flavour structures, including flavour anarchy and $U(3)^3$ or $U(2)^3$ flavour symmetries in the strong sector. In models with ``wrong-chirality'' Yukawa couplings, we find a strong bound 
from the neutron electric dipole moment, irrespective of the flavour structure. In the case of flavour anarchy, we also find strong bounds from flavour-violating dipoles, while these constraints are mild in the flavour-symmetric models.
\end{abstract}

\newpage

\section{Introduction}

The discovery of the Higgs boson \cite{Aad:2012tfa,Chatrchyan:2012ufa} has made the question what stabilizes the electroweak scale more acute and has reduced the implementation of weak-scale naturalness to two possibilities: supersymmetry and Higgs compositeness. In this paper, we study the latter possibility, which arguably has received less attention in the literature. This is in part due to the difficulty in treating the strong interactions that are responsible for the Higgs bound state. Much progress has been made in recent years by warped compactifications of higher-dimensional space-times \cite{Randall:1999ee,Contino:2003ve,Agashe:2004rs}, providing a weakly coupled dual description of the strong interactions in four dimensions. On the other hand, purely four-dimensional models have been constructed as well \cite{Contino:2006nn,Barbieri:2007bh,DeCurtis:2011yx,Panico:2011pw,Marzocca:2012zn}, with a particularly well-motivated example being models in which the Higgs arises as a pseudo Nambu-Goldstone 
boson, explaining its lightness with respect to the other, as yet unobserved, resonances\footnote{See also \cite{Bellazzini:2014yua} for a recent review of the composite Higgs model landscape in the light of the Higgs discovery.}.

In all these models, indirect constraints from low-energy precision observables play a crucial role. Generating fermion masses without excessive flavour violation singles out the mechanism of partial compositeness where, from a four-dimensional (4D) effective theory point of view, the elementary Standard Model (SM) fermions obtain masses by mixing linearly with composite vector-like fermion resonances \cite{Kaplan:1991dc}. Since the degree of compositeness is required to be smaller for light quarks, tree-level flavour-changing neutral currents (FCNCs) mediated by composite resonances are automatically suppressed \cite{Grossman:1999ra,Huber:2000ie,Gherghetta:2000qt}. However, unless one is willing to accept a fine-tuning of a few per cent, for TeV-scale resonances this suppression mechanism is not quite strong enough to suppress CP violation in kaon mixing, if no additional flavour symmetry is assumed \cite{Agashe:2004cp,Csaki:2008zd,Blanke:2008zb,Bauer:2009cf}. In addition to flavour constraints, the mixing 
of the SM fermions with composite states with different electroweak quantum numbers leads to potentially large corrections to electroweak precision observables. In particular, custodial symmetry should be imposed on the strong sector to avoid a tree-level correction to the $T$ parameter, and the representations of the composite fermions have to be chosen to avoid large tree-level corrections to the $Z\bar b_Lb_L$ vertex \cite{Agashe:2003zs,Agashe:2006at}.

In addition to electroweak precision observables and tree-level flavour-changing processes, an important class of constraints on these models is given by loop-induced dipole operators that contribute to radiative FCNC decays or fermionic dipole moments. The presence of heavy vector-like fermions charged under the electroweak gauge group implies a potential enhancement of these chirality-violating operators compared to the SM. A number of studies of dipole operators have been presented in the literature, either for Randall-Sundrum models in the KK basis \cite{Agashe:2004ay,Agashe:2004cp,Agashe:2008uz,Gedalia:2009ws,Delaunay:2012cz},
for purely 4D models \cite{Vignaroli:2012si}, or genuine five-dimensional (5D) calculations 
\cite{Csaki:2010aj,Blanke:2012tv,Beneke:2012ie}. The aim of this work is to exploit the computational simplicity of the 4D models to study the impact of different choices for the fermion representations and of different flavour symmetries on the constraints from observables sensitive to dipole operators. To this end, we will use a generalization of the framework of ref.~\cite{Contino:2006nn}, considering one set of composite partners for each SM field, plus the additional states required for custodial protection of $T$ and $Z\to b_L\bar b_L$.
Our work can be seen as a complement to similar studies of electroweak and tree-level flavour constraints \cite{Barbieri:2012tu} and $Z$-mediated rare decays \cite{Straub:2013zca} in the same models.

The remainder of this paper is organized as follows. In section~\ref{sec:setup} we define our model setup. Section~\ref{sec:pheno} is devoted to a model-independent discussion of dipole operators, the observables probing them and the constraints obtained from existing measurements. In section~\ref{sec:WilsonAnalytic} we derive approximate analytical expressions for the leading contributions to the Wilson coefficients of the dipole operators within our setup and estimate the size of subleading contributions. These results are then used in section~\ref{sec:numerics} to obtain numerical bounds on the masses of composite resonances derived from the various observables sensitive to dipole operators. While depending weakly on the choice of fermion representations, these bounds will depend strongly on the presence or not of a flavour symmetry in the strong sector. We will discuss the explicit examples of a $U(3)^3$ or a $U(2)^3$ flavour symmetry, 
only broken by the left- or right-handed composite-elementary mixings \cite{Barbieri:2008zt,Cacciapaglia:2007fw,Redi:2011zi,Barbieri:2012uh}. Since some results in the literature partially overlap with our results, it is mandatory that we compare our findings to them; we do this in section~\ref{sec:comp}. Section~\ref{sec:conc} contains our conclusions.

\section{Setup}
\label{sec:setup}

We consider simple four-dimensional models, in which partial compositeness is implemented along the lines of ref.~\cite{Contino:2006nn}: the SM field content (without the Higgs) is complemented by a composite Higgs, a set of vector resonances transforming under the global symmetry $G_c=SU(3)_c\times SU(2)_L\times SU(2)_R\times U(1)_X$ and a set of fermion resonances that fill complete representations of $G_c$. SM fermions and gauge bosons obtain masses from linear mixing terms with the composite states. Since $G_c$ is larger than the SM gauge group, there is some freedom in the choice of the fermion representations. In addition to the simplest case, involving just one $SU(2)_L$ doublet and one $SU(2)_R$ doublet (``doublet model''), we consider two cases (``triplet model'', ``bidoublet model'') that are motivated by the custodial protection of the $Zb\bar b$ coupling.

Using a notation where lower-case letters refer to elementary fields while upper-case letters denote composite states, the part of the Lagrangian involving fermions reads:
\begin{itemize}
\item In the doublet model,
\begin{gather}\label{doubletL}
\mathcal{L}^{\mathrm{doublet}}_s = 
- \bar Q^i_a m_Q^i Q^i_a - \bar R ^i_a m_R^i R^i_a
- \left(Y^{ij}  \bar Q^i_{L\,a} \mathcal H_{ab} R_{R\,b}^j
+ \tilde{Y}^{ij} \bar{R}_{L\,a}^i \mathcal H_{ba}^* Q_{R\,b}^j +\mathrm{h.c.}\right)\,, \\
\mathcal{L}^{\mathrm{doublet}}_{\mathrm{mix}} = 
\lambda^{ij}_L \bar q^i_{L\,a} Q^j_{R\,a} 
+ \lambda_{Ru}^{ij} \bar U^i_L t^j_R 
+ \lambda_{Rd}^{ij} \bar D^i_L b^j_R \,,
\end{gather}
where $\mathcal H=(i\sigma_2H^*,H)$ is the Higgs bidoublet, $Q = (T~B)^T$ transforms as a $(\mathbf{2},\mathbf{1})_{1/6}$ under $SU(2)_L\times SU(2)_R \times U(1)_X$, and $R=(U~D)$ transforms as a $(\mathbf{1},\mathbf{2})_{1/6}$.
Here and in the following, $i,j$ are flavour indices and $a,b,c$ are $SU(2)_L$ or $SU(2)_R$ indices.

\item In the triplet model,
\begin{align}
\mathcal{L}^{\mathrm{triplet}}_s = &
\bar L^i_{ab} m_{Q_u}^i L^i_{ba}
- \bar R^i_a m_R^i R^i_a - 
\bar R^{\prime i}_a m_R^i R^{\prime i}_a\nonumber \\
&-\left[
Y^{ij} \left(\bar L^i_{L}\right)_{ab} \mathcal H_{bc} \left(R^j_{R}\right)_{ca}+
Y^{ij} \left(\bar L^i_{L}\right)_{ab} \mathcal H_{bc} \left(R^{\prime j}_{R}\right)_{ca}+
\right.\nonumber \\
&+\left.
\tilde{Y}^{ij} \left(\bar R^i_{L}\right)_{ab} \mathcal H_{cb}^* \left(L^j_{R}\right)_{ca}+
\tilde{Y}^{ij} \left(\bar R^{\prime i}_{L}\right)_{ab} \mathcal H_{cb}^* \left(L^j_{R}\right)_{ca}+
\mathrm{h.c.}\right] \,, \label{tripletL} \\
\mathcal{L}^{\mathrm{triplet}}_{\mathrm{mix}} = &\,\,\lambda_L^{ij} \bar q_{L\,a}^i Q_{R\,a}^j + \lambda_{Ru}^{ij} \bar U^i_L t^j_R + \lambda_{Rd}^{ij} \bar D^i_L b^j_R \,,
\end{align}
where $L$ is a bidoublet transforming as a $(\mathbf{2},\mathbf{2})_{2/3}$, and $R$ and $R'$ are $SU(2)_R$ and $SU(2)_L$ triplets, transforming as $(\mathbf{1},\mathbf{3})_{2/3}$ and $(\mathbf{3},\mathbf{1})_{2/3}$, respectively. In component notation, the multiplets are given by
\begin{align}
L = (Q ~ Q') = \begin{pmatrix}
T & T_{5/3} \\ B & T_{2/3}
\end{pmatrix} \,, \quad R = (U_{5/3} ~ U ~ D)^T \,, \quad R' = (U'_{5/3} ~ U' ~ D') \,.
\end{align}
In the Yukawa couplings, we have also used
the triplets rewritten as $2\times 2$ matrices,
$R^{(\prime)}_{ab}=\tau^\alpha_{ab} R^{(\prime)}_\alpha$
with $\tau^{1,2}=(\sigma^1\pm i\sigma^2)/2$ and $\tau^3=\sigma^3/\sqrt{2}$.
\item In the bidoublet model
\begin{align}
\mathcal L_s^\text{bidoublet} =\
&-
\left(\bar L_{U}^i\right)_{ab}^i m_{Q_u}^i \left(L_{U}^i\right)_{ba}
-\bar U^i m_{U}^i U^i
\notag \\
 &
+\left[
Y_U^{ij} \,\left(\bar L_{U,L}^i\right)_{ab} \mathcal H_{ba} U_R^j
+\tilde{Y}_U^{ij}\, \bar U_L^i  \mathcal H_{ba}^* \left(L_{U,R}^j\right)_{ab} 
+\text{h.c} \right]
+(U\to D)
\,,\label{bidoubletL}
\\
\mathcal L_\text{mix}^\text{bidoublet} =
&\,\,\lambda_{Lu}^{ij}\bar q_{L\,a}^i Q_{Ru\,a}^j
+
\lambda_{Ru}^{ij}\bar U_L^i u_{R}^j
+ (U,u\to D,d)
\,,
\end{align}
where $L_U$ transforms as a $(\mathbf{2},\mathbf{2})_{2/3}$ and $L_D$ transforms as a $(\mathbf{2},\mathbf{2})_{-1/3}$ under the composite gauge group. $U$ and $D$ are singlets with the $U(1)_X$ charge $2/3$ and $-1/3$, respectively. The components of the multiplets are
\begin{align}
L_U = (Q_u~Q'_u) = \begin{pmatrix}
T & T_{5/3}\\ B & T_{2/3}
\end{pmatrix} \,, \quad L_D = (Q'_d~Q_d) = \begin{pmatrix}
B_{-1/3} & T' \\ B_{-4/3} & B'
\end{pmatrix}\,.
\end{align}
\end{itemize}

After rotating to the mass basis, the light and mostly elementary SM fermions couple to the Higgs through their mixings $\lambda$ with the composite states. For example, in the doublet model, the mass matrix of light quarks, after removing the mixing with the heavy fermions but before rotating to the mass basis, can be written as
\begin{align}
(m_{u,d})_{ij} &=
\frac{v}{\sqrt{2}}
\left(
\lambda_L
m_Q^{-1}
Y
m_R^{-1}
\lambda_{Ru,d}
\right)_{ij}
+
O\left(\frac{v^3}{m_{Q,R}^3}\right)
\,,
\label{eq:mass}
\end{align}
where $v=246$ GeV is the Higgs vacuum expectation value, and similar expressions hold in the bidoublet and triplet models. At leading order in $v/m_{Q,R}$, only the Yukawa couplings $Y$ (and not $\tilde Y$) enter the mass matrix, which is why the latter are sometimes called ``wrong-chirality'' Yukawa couplings. Although they are not necessary for the generation of quark masses, they are present in many models, and we will see that they play a crucial role in the generation of dipole operators, so we keep them in our Lagrangians.

\section{Model-independent phenomenology of dipole operators}\label{sec:pheno}

\subsection{Effective Hamiltonian}

We are interested in the electromagnetic and chromomagnetic dipole operators involving quarks, both flavour violating and flavour conserving. The relevant 
effective Hamiltonian can be written as
\begin{align}
\mathcal H_\text{eff} &= - \sum_{i,j,q,V}  C_{q_iq_jV} \, Q_{q_iq_jV} + C_{q_iq_jV}' \, Q_{q_iq_jV}'
\,,
\label{eq:Heff}
\end{align}
where $q=u,d$ and $V=\gamma,g$.
We define the dipole operators as
\begin{align}
Q_{q_iq_j\gamma} &= \frac{e\,m_{q_i}}{16 \pi^2} \,
 (\bar q_j \sigma^{\mu\nu} P_{R} q_i) \, F_{\mu\nu} \,,
&
Q_{q_iq_jg} &= \frac{g_s\,m_{q_i}}{16 \pi^2} \,
 (\bar q_j \sigma^{\mu\nu} T^a P_{R} q_i) \, G_{a\,\mu\nu} \,,
\,
\label{eq:O1}
\\
Q_{q_iq_j\gamma}' &= \frac{e\,m_{q_i}}{16 \pi^2} \,
 (\bar q_j \sigma^{\mu\nu} P_{L} q_i) \, F_{\mu\nu} \,,
&
Q_{q_iq_jg}' &= \frac{g_s\,m_{q_i}}{16 \pi^2} \,
 (\bar q_j \sigma^{\mu\nu} T^a P_{L} q_i) \, G_{a\,\mu\nu} \,.
\,
\label{eq:O2}
\end{align}
In the flavour-conserving case, one has $C'_{qqV}=C^*_{qqV}$, so in total there are 18 magnetic and 18 chromomagnetic quark dipole operators. Among those, the most phenomenologically relevant ones are the first-generation flavour-conserving operators contributing to the neutron EDM, the flavour-violating ones with down-type quarks contributing to FCNCs with $B$ and $K$ mesons, as well as $Q_{cuV}^{(\prime)}$ relevant for charm FCNCs. Before discussing the observables probing these operators in turn, we briefly summarize the QCD evolution that is necessary to relate the operators generated at a high new physics scale to the low-energy observables.

\subsection{QCD corrections}

The operators $O_{q_iq_j\gamma}^{(\prime)}$ and $O_{q_iq_j g}^{(\prime)}$ are subject to QCD renormalization and undergo mixing.
They evolve according to (omitting flavour indices)
\begin{equation}
\begin{pmatrix}
C_{\gamma}(\mu_l) \\ C_{g}(\mu_l)
\end{pmatrix}
=
\begin{pmatrix}
\eta_{\gamma\gamma} & \eta_{\gamma g} \\
0 & \eta_{gg}
\end{pmatrix}
\begin{pmatrix}
C_{\gamma}(\mu_h) \\ C_{g}(\mu_h)
\end{pmatrix}
,
\label{eq:rge}
\end{equation}
and equivalently for the primed coefficients. For the running from some high new physics matching scale $\mu_h$ down to the top mass $m_t$, one has at leading logarithmic order \cite{Buchalla:1995vs}
\begin{align}
\eta_{\gamma\gamma} &= \left[\frac{\alpha_s(\mu_h)}{\alpha_s(m_t)}\right]^{16/21}
\,,&
\eta_{\gamma g} &= \frac{8}{3}\left(
\left[\frac{\alpha_s(\mu_h)}{\alpha_s(m_t)}\right]^{2/3}
-
\left[\frac{\alpha_s(\mu_h)}{\alpha_s(m_t)}\right]^{16/21}
\right)
\,,&
\eta_{gg} &= \left[\frac{\alpha_s(\mu_h)}{\alpha_s(m_t)}\right]^{2/3}
\,.
\end{align}
For the evolution from $m_t$ down to some low scale $\mu_l$, the number of active quark flavours change and quark mass thresholds have to be taken into account. We list numerical values of the evolution coefficients $\eta$ for the evolution from some exemplary high scale values to $m_t$, as well as from $m_t$ to phenomenologically relevant low scales, in table~\ref{tab:eta}.
\begin{table}
\centering
\renewcommand{\arraystretch}{1.1}
\begin{tabular}{cccc}
\hline
$\mu_h$ &  $\eta_{\gamma\gamma}$ & $\eta_{\gamma g}$ & $\eta_{gg}$ \\
\hline
$0.5$ TeV & $0.905$ & $0.030$ & $0.917$ \\
$1$ TeV & $0.856$ & $0.045$ & $0.873$ \\
$2$ TeV & $0.813$ & $0.057$ & $0.835$ \\
$5$ TeV & $0.763$ & $0.070$ & $0.789$ \\
\hline
\end{tabular}
\qquad
\begin{tabular}{cccc}
\hline
$\mu_l$ &  $\eta_{\gamma\gamma}$ & $\eta_{\gamma g}$ & $\eta_{gg}$ \\
\hline
$m_W$ & $0.930$ & $0.023$ & $0.939$ \\
$m_b$ & $0.603$ & $0.105$ & $0.642$ \\
$2\,\text{GeV}$ & $0.502$ & $0.120$ & $0.547$ \\
$m_c$ & $0.432$ & $0.127$ & $0.480$ \\
$1\,\text{GeV}$ & $0.389$ & $0.130$ & $0.438$ \\
\hline
\end{tabular}
\caption{RG coefficients for the evolution from some high new physics scale $\mu_h$ to $m_t$ (left), and from $m_t$ to some low energy scale $\mu_l$. We use $\alpha_s(M_Z)=0.1185$.}
\label{tab:eta}
\end{table}

In writing eq.~(\ref{eq:rge}), we have neglected the mixing of neutral or charged current-current (four-quark) operators into the dipole operators \cite{Buras:2011zb}. Although such operators are generated at tree-level in our setup, we expect that their contributions to the dipole operators are small, since they are suppressed by additional powers of the composite-elementary mixing angles.

\subsection{Neutron EDM}

The electric and chromoelectric dipole moments (EDMs and CEDMs) of the quarks are related to the Wilson coefficients of the flavour-conserving dipole operators as
\begin{align}
d_{q} = \frac{e \, m_q}{8\pi^2} \ \mathrm{Im}\left( C_{qq\gamma}(\mu_l) \right)
\,,
\qquad
\tilde d_q = \frac{g_s \, m_q}{8\pi^2} \ \mathrm{Im}\left( C_{qqg}(\mu_l) \right)
\,, \label{eq:quarkEDMs}
\end{align}
where $\mu_l$ is a hadronic scale of order 1~GeV. The calculation of the contributions of the quark (C)EDMs to the neutron EDM is plagued by considerable hadronic uncertainties. An estimate obtained using QCD sum rules \cite{Pospelov:2000bw} yields
\begin{equation} \label{neutron}
d_n = (1 \pm 0.5) \left[ 1.4\left( d_d - \tfrac{1}{4} d_u\right) + 1.1 e \left( \tilde{d}_d + \tfrac{1}{2} \tilde{d}_u\right)  \right] .
\end{equation}
Experimentally, the neutron EDM is already strongly constrained \cite{Baker:2006ts},
\begin{equation}
|d_n| < 2.9 \times 10^{-26} ~e\,\text{cm}  ~~\text{at 90\% C.L.}
\label{eq:dnexp}
\end{equation}
Several experiments are in construction that plan to improve this bound by up to two orders of magnitude \cite{Hewett:2012ns}.

Indirectly, the neutron EDM is also sensitive to the CEDMs of second and third generation quarks. In the QCD evolution of the CEDMs to low energies, when integrating out a heavy quark, a finite threshold correction is generated to the three-gluon Weinberg operator, which directly contributes to the neutron EDM and mixes under renormalization with the first generation quark (C)EDMs \cite{Braaten:1990gq}. Taking these effects into account, the bound (\ref{eq:dnexp}) can be translated into bounds on the charm, bottom and top CEDMs, which read \cite{Chang:1990jv,Kamenik:2011dk,Sala:2013osa}
\begin{align}
|\tilde d_c|&<1.0\times10^{-22} \,\text{cm} \,,
&
|\tilde d_b|&<1.1\times10^{-21} \,\text{cm} \,,
&
|\tilde d_t|&<2.1\times10^{-19} \,\text{cm} \,.
\end{align}

\subsection{Down-type FCNCs}

The most well-measured flavour-changing dipole transitions are the $b\to s\gamma/g$ processes probed in the inclusive decay $B\to X_s\gamma$. The corresponding decay probing the $b\to d\gamma/g$ transitions is even rarer in the SM due to the stronger CKM suppresion and consequently is measured less precisely. Normalizing the current experimental measurements to the SM expectations for the branching ratios,
\begin{equation}
R_{bq\gamma} = \frac{\text{BR}({B} \to X_q \gamma)}{\text{BR}({B} \to X_q \gamma)_\text{SM}}
\,, \quad \mbox{with~ $q=s,d$,}
\label{eq:Rbqg}
\end{equation}
one has at present \cite{Amhis:2012bh,Misiak:2006zs,Becher:2006pu,Crivellin:2011ba}
\begin{equation}
R_{bs\gamma}=1.13\pm0.11
\,,\qquad
R_{bd\gamma}=0.92\pm0.40
\,.
\label{eq:bqgexp}
\end{equation}
Beyond the SM, these quantities are modified as \cite{Kagan:1998ym,Buras:2011zb}
\begin{equation}
R_{bq\gamma} = 1 + 0.97 \left( 2\,\text{Re}(R_{7q}) + |R_{7q}|^2 + |R_{7q}'|^2 \right),
\end{equation}
where
\begin{equation}
R_{7q}^{(\prime)} = \frac{\sqrt{2}}{4 G_F V_{tb}V_{tq}^*} \frac{C^{(\prime)}_{bq\gamma}(m_b)}{C_7^\text{eff}(m_b)}\,
\,,
\end{equation}
with $C_7^\text{eff}(m_b)=-0.3523$. For the numerical bounds on the Wilson coefficients in the next sections, we imposed the constraints (\ref{eq:bqgexp}) at $2\sigma$.

The $s\to d\gamma/g$ transitions are less constrained experimentally, since the long-distance dominance in $K$ decay processes makes it difficult to relate experimental observables to the short-distance contributions. Nevertheless, a meaningful bound on the Wilson coefficients $C_{sdg}^{(\prime)}$ can be obtained from the measurement of the parameter $\epsilon'/\epsilon$. With the conservative assumption that the new physics contribution to $\epsilon'/\epsilon$ should not exceed its experimental central value, one obtains the bound \cite{Mertens:2011ts}
\begin{equation}
\frac{1}{2}\,\text{Im} \left(C_{sdg}-C_{sdg}'\right) < 3.1 \times 10^{-8}
\,.
\end{equation}

\subsection{Charm FCNCs}

Recent experimental hints that the direct CP asymmetry difference $\Delta A_\text{CP}$ between $D\to KK$ and $D\to \pi\pi$ decays is larger than the SM expectation have attracted a lot of interest as a possible sign of new physics, also in the context of models with partial compositeness \cite{KerenZur:2012fr,Delaunay:2012cz}. But even if the observed effect is not due to new physics, the upper bound on $\Delta A_\text{CP}$ can be used to put a constraint on the charm chromomagnetic dipole operator $Q'_{cug}$ \cite{Isidori:2011qw}. Following \cite{Sala:2013osa}, we impose in the numerical analysis that the new physics contribution to $\Delta A_\text{CP}$, for central values of the hadronic parameters, does not exceed the world average \cite{Amhis:2012bh}
\begin{equation}
\Delta A_\text{CP} = (-0.319\pm0.121)\% \,.
\end{equation}

\subsection{Model-independent bounds}

Given all the experimental constraints discussed above, we can derive model-independent bounds on the Wilson coefficients of the dipole operators. We list them in table~\ref{tab:mibounds} at a renormalization scale of 1~TeV, considering one purely real or purely imaginary Wilson coefficient at a time.

The only operators in the effective Hamiltonian (\ref{eq:Heff}) we have not considered are
the flavour-changing ones involving top quarks. Although they are  not yet strongly constrained,
they will be probed at LHC in the future through the decays $t\to q\gamma$ and $t\to qg$, where $q=u,c$.

\begin{table}[t]
\renewcommand{\arraystretch}{1.4}
\centering
\begin{tabular}{cccccc}
\hline
operator & $\text{Re}(C)<M^{-2}$ & $\text{Re}(C)>-M^{-2}$ & $\text{Im}(C)<M^{-2}$ & $\text{Im}(C)>-M^{-2}$  & process \\
\hline
$Q_{uu\gamma}$ && & \multicolumn{2}{c}{$1.08$ TeV}&  $d_n$ \\
$Q_{dd\gamma}$ && & \multicolumn{2}{c}{$3.11$ TeV}&  $d_n$ \\
$Q_{uug}$ && & \multicolumn{2}{c}{$1.45$ TeV}&  $d_n$ \\
$Q_{ddg}$ && & \multicolumn{2}{c}{$3.79$ TeV}&  $d_n$ \\
$Q_{ccg}$ && & \multicolumn{2}{c}{$1.22$ TeV}&  $d_n$ \\
$Q_{bbg}$ && & \multicolumn{2}{c}{$0.67$ TeV}&  $d_n$ \\
$Q_{ttg}$ && & \multicolumn{2}{c}{$0.30$ TeV}&  $d_n$ \\
\hline
$Q_{bs\gamma}$ &$0.71$ TeV& $2.81$ TeV& $1.44$ TeV& $1.39$ TeV & $B\to X_s\gamma$ \\
$Q_{bsg}$ &$0.34$ TeV &$1.34$ TeV& $0.69$ TeV& $0.67$ TeV& $B\to X_s\gamma$ \\
$Q'_{bs\gamma}$ &\multicolumn{2}{c}{$1.41$ TeV} &\multicolumn{2}{c}{$1.31$ TeV}  & $B\to X_s\gamma$ \\
$Q'_{bsg}$ &\multicolumn{2}{c}{$0.68$ TeV}&\multicolumn{2}{c}{$0.68$ TeV}  & $B\to X_s\gamma$ \\
\hline
$Q_{bd\gamma}$ &$3.74$ TeV& $1.51$ TeV& $2.91$ TeV& $1.94$ TeV & $B\to X_d\gamma$ \\
$Q_{bdg}$ &$1.79$ TeV &$0.72$ TeV& $1.40$ TeV& $0.93$ TeV& $B\to X_d\gamma$ \\
$Q'_{bd\gamma}$ &\multicolumn{2}{c}{$2.37$ TeV} &\multicolumn{2}{c}{$2.37$ TeV}  & $B\to X_d\gamma$ \\
$Q'_{bdg}$ &\multicolumn{2}{c}{$1.14$ TeV}&\multicolumn{2}{c}{$1.14$ TeV}  & $B\to X_d\gamma$ \\
\hline
$Q^{(\prime)}_{sdg}$ &&& \multicolumn{2}{c}{$2.80$ TeV}& $\epsilon'/\epsilon$ \\
\hline
$Q^{(\prime)}_{cug}$ &&&\multicolumn{2}{c}{$2.14$ TeV}& $D\to KK,\pi\pi$ \\
\hline
\end{tabular}
\caption{Model-independent bounds on new physics contributions to Wilson coefficients of dipole operators. The four columns show the lower bounds on $M$, where the Wilson coefficients at the matching scale of 1~TeV were assumed to be  $C_{q_iq_jV}(1\,\text{TeV})=(1,-1,i,-i)/M^2$.}
\label{tab:mibounds}
\end{table}

\section{Analytical results for the Wilson coefficients}\label{sec:WilsonAnalytic}

In this section, we derive approximate analytical expressions for the Wilson coefficients of the dipole operators for the three different choices of fermion representations. We denote by 
$M\sim m_{Q,R}$ a generic composite mass, by $\lambda$ a generic composite-elementary mixing parameter,  by $g$ an elementary gauge coupling and by $g_\rho$ the coupling of the composite vector resonances. Our goal is to obtain expressions for the Wilson coefficient to a given order in the small ratios $v/M$, $\lambda/M$, and $g/g_\rho$. %
To this end, we first consider the case of a single generation of elementary and composite fermions. The relevant mass matrices arising in the three models are collected in appendix~\ref{app:massMatrices}. We diagonalize these matrices at a given order in the small ratios, rotate all couplings to the mass eigenstate basis, and compute the Wilson coefficients. The resulting one-loop expressions for the contributions to the Wilson coefficients involving scalars or vectors are listed along with the relevant loop functions in appendix~\ref{app:reference}. It turns out that the dominant contributions typically arise from diagrams with a heavy fermion -- lifting the chirality suppression -- together with a $W$, $Z$ or Higgs in the loop. In section~\ref{sec:leading}, we first discuss these contributions in detail, before qualitatively discussing the additional contributions in section~\ref{sec:subleading}.

\subsection{Leading contributions}
\label{sec:leading}

For a single generation of fermions, to leading order in the small parameters $v/M$, $\lambda/M$ and in the limit of heavy vector resonances, we find that the Wilson coefficients can be written in the form\footnote{%
Here and in the following, to simplify the notation, we will assume $m_{Q_u}=m_{Q_d}\equiv m_Q$, $m_U = m_D \equiv m_R$, $Y_{U}=Y_{D}\equiv Y$, and $\tilde Y_{U}=\tilde Y_{D}\equiv \tilde Y$ in the bidoublet model.
}
\begin{align}
C_{qqV} &=
C_{qqV}^\text{SM}
+
a_{qV} \frac{Y \tilde Y}{m_Q m_R}
\,,
\label{eq:CLO}
\end{align}
where $q=u,d$ and $V=\gamma,g$. In this limit, the only relevant diagrams feature a Higgs, $W$ or $Z$ boson as well as a heavy fermion in the loop. We have computed all coefficients $a_{qV}$ in the doublet, triplet, and bidoublet models and list them in table~\ref{tab:a12}. To illustrate our procedure, we give a detailed account of our calculation of $a_{d\gamma}$ in the bidoublet model in appendix~\ref{app:example}. We note that the value $a_{u\gamma}=0$ in the doublet model is not due to a symmetry but rather due to an accidental cancellation between the $W$, $Z$ and Higgs contributions.

\begin{table}
\renewcommand{\arraystretch}{1.4}
\centering
\begin{tabular}{cccc}
\hline
& doublet & triplet & bidoublet \\
\hline
$a_{d\gamma}$ & $ \frac{1}{4} $ & $ \frac{1}{2} $ & $ -\frac{1}{2} $\\
$a_{dg}$ & $\frac{3}{4} $ & $ \frac{3}{2} $ &$ \frac{3}{2} $ \\
\hline
\end{tabular}
\qquad
\begin{tabular}{cccc}
\hline
& doublet & triplet & bidoublet \\
\hline
$a_{u\gamma}$ & $ 0 $ & $ \frac{10}{3} $ &  $ 1 $ \\
$a_{ug}$ &  $ \frac{3}{4} $ &  $ 2 $ & $ \frac{3}{2} $ \\
\hline
\end{tabular}
\caption{Coefficients entering the leading-order contribution (\ref{eq:CLO}) to the dipole Wilson coefficients of down-type quarks (left) and up-type quarks (right).}
\label{tab:a12}
\end{table} 

An important result of our calculation is that, at leading order, there is no quadratic term in $Y$, as was also emphasized in ref.~\cite{Delaunay:2012cz} in the context of the $c\to ug$ dipole transition\footnote{The proportionality to the wrong-chirality Yukawa was also found for the $\mu\to e\gamma$ dipole in ref.~\cite{Agashe:2006iy}.}. This means that in models in which the ``wrong-chirality'' Yukawa couplings are absent or suppressed, the dipole operators will be suppressed as well. We will discuss other contributions, that become the leading ones in the limit $\tilde Y\to 0$, in section~\ref{sec:subleading}.

The result in eq.~(\ref{eq:CLO}) is only valid in the unrealistic case of a single generation of fermions. Taking into account all three generations and an arbitrary flavour structure,
the full analytic diagonalization of the mass matrices is clearly not feasible.
Still, it is possible to obtain an approximate analytical expression valid for three generations of elementary and composite quarks by promoting eq.~(\ref{eq:CLO}) to a matrix equation in flavour space. Concretely, for $i\geq j$, one has 
\begin{align}
C_{d_id_jV} &= C_{d_id_jV}^\text{SM} + \frac{a_{dV}}{m_{d_i}} \Delta^d_{ji}
\,,
\label{eq:CLO31}
\\
C_{d_id_jV}' &= \frac{a_{dV}}{m_{d_i}} \Delta^d_{ij}
\,,
\\
\text{where}~~
\Delta^d &= \frac{v}{\sqrt2}\, U_{Ld}^\dagger \lambda_L m_Q^{-1} Y m_R^{-1} \tilde Y
 m_Q^{-1} Y m_R^{-1} \lambda_{Rd} U_{Rd}
\,,
\label{eq:CLO32}
\end{align}
and analogously for up-type quarks\footnote{In the triplet model, there is an additional factor of $\frac{1}{\sqrt{2}}$ in front of $\Delta^u$ compared to eqs.~(\ref{eq:CLO31})--(\ref{eq:CLO32}), cf.\ the mass matrices in appendix~\ref{app:massMatrices}.}, where $U_{Lq,Rq}$ are the matrices diagonalizing the quark mass terms (\ref{eq:mass}). We checked numerically that eqs.~(\ref{eq:CLO31})--(\ref{eq:CLO32}) indeed give a very good approximation to the exact results obtained by numerically diagonalizing the mass matrices.

\subsection{Subleading contributions}
\label{sec:subleading}

Going beyond the leading order in the expansion of composite-elementary mixings and $v/M$ and beyond the limit of heavy vector resonances, there are several classes of contributions that can become relevant in some cases, in particular in models without wrong-chirality Yukawa couplings. In general, no simple analytical expressions can be given for these subleading contributions, so our discussion will remain qualitative.

\subsubsection{Higher orders in elementary-composite mixing}
\label{sec:NLO1}

These contributions are suppressed by $\lambda^2/M^2$ with respect to (\ref{eq:CLO}) and are relevant for Wilson coefficients involving the third generation, that can have a sizable degree of compositeness, in particular for $b\to q\gamma/g$. These contributions can arise
\begin{enumerate}[a)]
\item from diagrams with a SM quark and a $W$ or $Z$ in the loop and an $O(v^2/M^2)$ correction to the quark-gauge boson vertex;
\item from diagrams with a heavy fermion and a $W$, $Z$, or Higgs in the loop that are parametrically of the same order.
\end{enumerate}
We start by discussing the contributions of type a). In the case of $b\to s\gamma/g$ and $b\to d\gamma/g$, there are two contributions to the $W$--top loop that are only suppressed by the degree of compositeness of the right-handed top quark compared to the leading contribution. They read
\begin{align}
\delta C_{bqV} &= \frac{4G_F}{\sqrt{2}} \frac{m_t}{m_b} V_{tq}^* \,  (\delta g_W^R)_{tb} \,f_V(x_t)
\label{eq:IRL}
\,,\\
\delta C_{bqV}' &= \frac{4G_F}{\sqrt{2}} \frac{m_t}{m_b} V_{tb} \,  (\delta g_W^R)_{tq} \,f_V(x_t)
\,,
\label{eq:IRR}
\end{align}
where $q=d,s$, $x_t=m_t^2/m_W^2$, and the $W$ coupling of the SM quarks is written as
\begin{equation}
\mathcal L \supset \frac{g}{\sqrt{2}} \bar u_i
\left[
\left(V_{ij}+(\delta g_W^L)_{ij}\right)\gamma^\mu P_L + (\delta g_W^R)_{ij}\gamma^\mu P_R
\right] d_j W_\mu^+ .
\end{equation}
In general, one has
\begin{align}
(\delta g_W^L)_{ij}&=a\,v^2 \, \left(\lambda_L m_Q^{-1} YY^\dagger m_Q^{-1} \lambda_L\right)_{ij}
\,,&
(\delta g_W^R)_{ij}&=b\,v^2 \, \left(\lambda_{Ru} m_R^{-1} Y^\dagger Y m_R^{-1} \lambda_{Rd}\right)_{ij}
\,,
\label{eq:Wc}
\end{align}
where the coefficients $a$ and $b$ are given in table \ref{tab:WCorrections}.
\begin{table}
\renewcommand{\arraystretch}{1.4}
\centering
\begin{tabular}{cccc}
\hline 
 & doublet & triplet & bidoublet \\ 
\hline 
$a$ & $-\frac{1}{2}$ & $-\frac{1}{4}$ & $-\frac{1}{4}$ \\ 
$b$ & $\frac{1}{2}$ & $\frac{\sqrt{2}}{4}$ & $0$ \\ 
\hline 
\end{tabular} 
\caption{Coefficients in eq.~(\ref{eq:Wc}) relevant for the corrections to the $W$ couplings in all models.}\label{tab:WCorrections}
\end{table}

The contribution in (\ref{eq:IRL}) due to the right-handed $Wtb$ coupling was first discussed in ref.~\cite{Vignaroli:2012si}. We emphasize that the contribution in (\ref{eq:IRR}) can be equally important in specific models, although it depends on the flavour structure of $\delta g_W^R$, while the contribution in (\ref{eq:IRL}) is present even for a flavour diagonal $\delta g_W^R$. Concerning the contributions of type b), which involve heavy fermions in the loop, we merely note that they are parameterically of the same order as the ones of type a), but can have a different flavour structure and are therefore more model dependent. Contributions involving the degree of compositeness of the left-handed bottom quark are suppressed by $m_b/m_t$, and we will negelect them.

\subsubsection{Higher orders in inverse powers of the composite mass scale}
\label{sec:Omh}

We now consider terms that do not involve additional composite-elementary mixings, but are present even in the limit $\tilde Y\to0$. These contributions are relevant for Wilson coefficients {\em not} involving the third generation in models where the ``wrong-chirality'' Yukawas $\tilde Y$ are absent or suppressed. Such contributions arise for example from an expansion of the loop functions of the diagrams with a $W$, $Z$ or Higgs and a heavy fermion to higher order in the ratios $x=m_\psi^2/m_{W,Z,h}^2$, where $m_\psi$ is the heavy fermion mass. In fact, the analytic cancellation of the contributions proportional to $Y^2$ works only at the leading order of the expansion of the loop functions. As an example, we discuss the Higgs contribution to the down-type quark dipole operator for a single generation in the bidoublet model for $\tilde Y=0$. We find
\begin{align}
C_{ddV} &\supset
-\frac{1}{6}
\frac{m_h^2 \left(m_Q^2+m_R^2\right)Y^2}{m_Q^3m_R^3}
\,.
\label{eq:CNLO2}
\end{align}
There are similar contributions suppressed by ${m_W^2}/{m_\psi^2}$ and ${m_Z^2}/{m_\psi^2}$. In the case of the $W$ contribution, there is the special feature that, in addition to the quadratic term in the mass ratio, there is also a logarithm that becomes dominant for large fermion resonance mass. In view of this complicated dependence, we refrain from giving full analytical expressions for this type of subleading contributions in all models, but simply keep in mind that, in the absence of wrong-chirality Yukawa couplings and sizable composite-elementary mixings, the Higgs, $W$ and $Z$ contributions to the dipole operators are roughly suppressed by $m_{h,W,Z}^2/m_\psi^2$ compared to the leading contribution for non-zero $\tilde Y$.

\subsubsection{Higher loop orders}

\begin{figure}
\centering
\includegraphics[width=5.5cm]{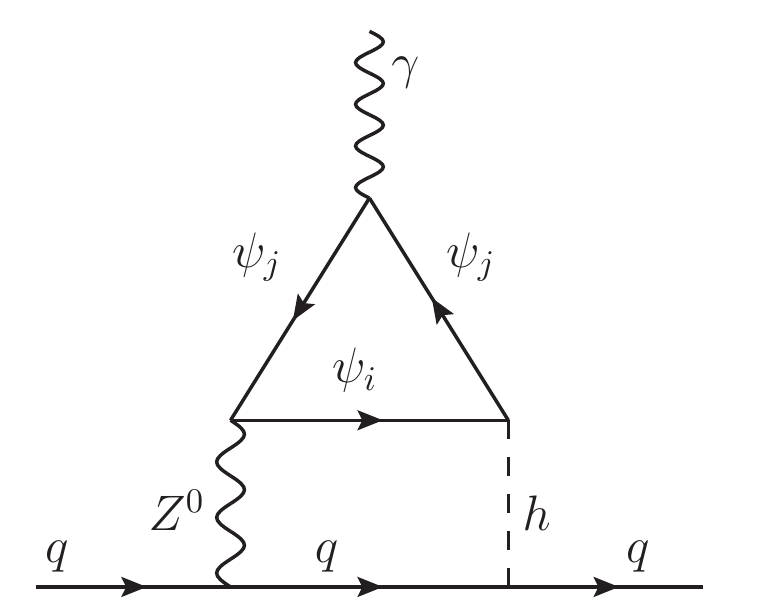}
\caption{Two-loop Barr-Zee type diagram contributing to the electromagnetic dipole operator.}
\label{fig:BarrZee}
\end{figure}

Two-loop contributions to the dipole operators might be relevant in cases where
the wrong-chirality Yukawas are absent or strongly suppressed,
operators {\em not\/} involving the third generation are considered (in particular, EDMs), 
and the composite mass scale is large. For operators involving the third generation, the contributions discussed in sec.~\ref{sec:NLO1} dominate instead.
The last item is relevant because the contributions discussed in section~\ref{sec:Omh} decouple with the fourth power of the inverse mass scale, while at two-loop order, there can be diagrams that decouple with the square of the inverse mass scale, but that do not vanish for $\tilde Y=0$. An example is given by the Barr-Zee type diagram shown in fig.~\ref{fig:BarrZee}, that is familiar from the two-loop chargino contribution to the EDM in split supersymmetry \cite{Giudice:2005rz}. We estimate the contribution of this diagram to the Wilson coefficient in the limit $\tilde Y=0$ as
\begin{equation}
C_{q_iq_jV} \sim \frac{g^2}{16\pi^2} \, \frac{Y^2}{m_\psi^2} \,,
\label{eq:CNLO3}
\end{equation}
up to an $O(1)$ factor. We see that it can be safely neglected with respect to the leading contribution (\ref{eq:CLO}) even for $\tilde Y\sim Y$, but it can dominate compared to the contribution (\ref{eq:CNLO2}) if $m_\psi\sim m_Q\sim m_R$ is in the multi-TeV regime.

\subsubsection{Diagrams with heavy vector resonances}

Until now, we have only considered one-loop diagrams with a heavy fermion and a $W$, $Z$ or Higgs in the loop, but there are also diagrams with a heavy vector resonance and a heavy fermion. These contributions are always parametrically suppressed by a factor
$g_\rho^2/m_\rho^2$, where $g_\rho$ is the coupling and $m_\rho$ the mass of the vector resonance. In general, the analytical expressions for these contributions are complicated, since, in contrast to the $W$ or $Z$ contributions considered above, one has to keep the full dependence of the loop functions if the fermion and vector resonance masses are comparable. However, it is important to notice that in the limit where all the fermion resonances are degenerate, the contribution to the dipole operators from these diagrams is real and diagonal in the mass basis and thus does not contribute to any of the observables we consider, which always feature either flavour or CP violation\footnote{In the lepton sector, which we do not consider here, the muon anomalous magnetic moment is an important exception, because it does not require any flavour or CP violation.}.
In general, we find that in the case of non-zero $\tilde Y$, these contributions are always suppressed by $v^2 g_\rho^2/m_\rho^2$ and/or $m_\psi^2/m_\rho^2$ (which is preferred to be smaller than~1 since naturalness prefers light fermion resonances and electroweak precision tests require heavy vector resonances) with respect to the leading contribution (\ref{eq:CLO}), and we confirmed with a numerical scan that they are typically small. We will not consider this class of contributions in the following, but one should keep in mind that, in particular corners of the parameter space, they might be relevant in specific models and would tighten the bounds considered below.

\subsubsection{Higher-dimensional operators}
\label{sec:NLO-1}

Finally, in a more complete theory like a composite Higgs model, there can be additional contributions that are not captured by our Lagrangians defined at the beginning of section~\ref{sec:setup} and are therefore not calculable in our setup. This means that the bounds we obtain below can be viewed as conservative estimates. It is possible that there are additional contributions that make the bounds more severe; but, on general grounds, there is no reason to expect that these additional effects conspire with the calculable ones to eliminate the constraints.

\section{Phenomenological analysis}
\label{sec:numerics}

We now proceed to a numerical analysis of the bounds on partial compositeness from observables sensitive to dipole operators. Since all these observables probe either flavour or CP violation (or both), the bounds crucially depend on the assumptions made on the flavour structure of the model. We start with the most popular assumption of flavour anarchy, which arises in models aiming at a geometrical explanation for the quark mass and mixing hierarchies but is known to have a problem (assuming TeV-scale resonance masses) with excessive CP violation in $K$ mixing, unless one is willing to accept an $O(10^{-2})$ fine tuning of the revelvant CP-violating phase. We then also consider models with a global flavour symmetry in the strong sector -- either $U(3)^3$ or $U(2)^3$ -- only broken by the composite-elementary mixings.

Our aim in this section is {\em not} to perform a full numerical analysis of these models and the contributions to dipole operators. Rather, we aim at providing approximate analytical expressions for the dominant contributions to the dipole operators and use them to extract approximate lower bounds on the resonance masses from the experimental measurements. These results can then be used to judge how severely a model with a given choice of fermion representations and with a given flavour structure is constrained by the observables sensitive to dipole operators.

\subsection{Models with flavour anarchy}

We first consider the case of flavour anarchy, where all the couplings in the strong sector are assumed to have $O(1)$ off-diagonal elements and phases.
In general, all coefficients then depend on complicated functions of the anarchic Yukawa and mass matrix elements. To give simplified approximate expressions one can use the fact that, up to $O(1)$ factors, the quark Yukawa couplings and the CKM matrix elements can be written in terms of the degrees of compositeness $s_{L,R}$ (see appendix~\ref{app:massMatrices} for their definition) as
\begin{align*}
y_{u_i} &\sim Y s_{Li} s_{Ru_i} \,,
&
y_{d_i} &\sim Y s_{Li} s_{Rd_i} \,,
&
V_{ij} & \sim s_{Li}/s_{Lj} ~~(j>i) \,.
\end{align*}
where $Y$ can be understood as an ``average'' Yukawa coupling. 
In the following, we provide simplified expressions for the Wilson coefficients in terms of ``average'' parameters $Y$, $\tilde Y$ and $m_i$ that keep track of how the quantities scale with the parameters, but we neglect $O(1)$ factors coming from the flavour structure. We do however take into account the numerical factors derived in section~\ref{sec:WilsonAnalytic}.
The leading contributions to the Wilson coefficients at the matching scale then read
\begin{align}
C_{qqV} &\sim a_{qV} \frac{Y \tilde{Y}}{m_Q m_R} \qquad\text{for }q=u,d
\,, \\
C_{bqV} &\sim V_{tq} \, a_{dV} \frac{Y \tilde{Y}}{m_Q m_R}
\qquad\text{for }q=d,s
\,,\\
C_{bqV}' &\sim \frac{m_q}{m_bV_{tq}}a_{dV} \frac{Y \tilde{Y}}{m_Q m_R}
\qquad\text{for }q=d,s
\,,\\
C_{sdg} &\sim  V_{cd} \,a_{dg} \frac{Y \tilde{Y}}{m_Q m_R}
\,, \\
C_{sdg}' &\sim  \frac{m_d}{m_sV_{cd}} a_{dg} \frac{Y \tilde{Y}}{m_Q m_R}
\,, \\
C_{cug} &\sim  V_{us} \,a_{ug} \frac{Y \tilde{Y}}{m_Q m_R}
\,,\\
C_{cug}' &\sim  \frac{m_u}{m_cV_{us}} a_{ug} \frac{Y \tilde{Y}}{m_Q m_R}
\,,
\end{align}
where $V=\gamma,g$. Arbitrary phases and $O(1)$ factors are understood in all cases. Concerning the relative importance of the primed and unprimed flavour-changing Wilson coefficients, it is interesting to note that in $b\to s$ and $b\to d$ transitions the flavour prefactor is an order of magnitude larger for the primed coefficients, so observables in $B$ decays sensitive to
the primed Wilson coefficients, i.e.\ to right-handed flavour-changing neutral currents, are particularly promising in the anarchic model (see \cite{Becirevic:2012dx} for an overview of promising observables). 
For the $s\to d$ transition, the prefactors of primed and unprimed coefficients are comparable, and for the $c\to u$ transition the unprimed coefficient has a prefactor that is about a factor 30 larger than the unprimed one.

For the $b\to s$ and $b\to d$ transitions, there is an additional important contribution that is only suppressed by the degree of compositeness of the right-handed top quark, as discussed in section~\ref{sec:subleading}. Here, we give only a crude parametrical estimate of this contribution,
\begin{align}
\delta C_{bqV} &\sim V_{tq} \frac{Y^2}{m_Q^2} s_{Rt}^2 
\,,\\
\delta C'_{bqV} &\sim \frac{m_q}{m_bV_{tq}} \frac{Y^2}{m_Q^2} s_{Rt}^2 
\,.
\end{align}
For all Wilson coefficients, there is in addition a subleading contribution not involving $\tilde Y$ that is parametrically suppressed by $m_{h,W,Z}^2/(m_Qm_R)$ compared to the leading one (for $\tilde Y\sim Y$), as discussed in section~\ref{sec:Omh}.

\begin{table}[tb]
\renewcommand{\arraystretch}{1.4}
\centering
\begin{tabular}{cccccc}
\hline
bound on:
& \multicolumn{3}{c}{$\left(\frac{m_Q m_R}{Y \tilde Y}\right)^{1/2}$}
& $\left(\frac{m_Q^2}{s_{Rt}^2 Y^2}\right)^{1/2}$
&$\left(\frac{m_Q m_R}{Y}\right)^{1/2}$
\\
operator & doublet & triplet & bidoublet & (estimate) & (estimate) \\
\hline
$Q_{ddV}$& $3.6$ TeV& $5.1$ TeV& $4.1$ TeV && $0.8$ TeV\\
$Q_{uuV}$& $1.3$ TeV& $0.6$ TeV& $1.4$ TeV && $0.3$ TeV\\
$Q_{ccg}$& $1.1$ TeV& $1.7$ TeV& $1.5$ TeV && $0.5$ TeV\\
$Q_{bbg}$& $0.6$ TeV& $0.9$ TeV& $0.8$ TeV && $0.3$ TeV\\
$Q_{ttg}$& $0.3$ TeV& $0.4$ TeV& $0.4$ TeV && $0.2$ TeV\\
$Q_{bsV}$& $0.4$ TeV& $0.5$ TeV& $0.2$ TeV  &$0.6$ TeV& $0.3$ TeV\\
$Q'_{bsV}$& $0.7$ TeV& $1.0$ TeV& $0.4$ TeV &$1.1$ TeV& $0.3$ TeV\\
$Q_{bdV}$& $0.2$ TeV&$0.3$ TeV&  $0.1$ TeV &$0.3$ TeV& $0.2$ TeV\\
$Q'_{bdV}$& $0.6$ TeV&  $0.8$ TeV& $0.3$ TeV &$0.9$ TeV& $0.3$ TeV\\
$Q_{sdg}$& $1.1$ TeV& $1.6$ TeV& $1.6$ TeV && $0.5$ TeV\\
$Q'_{sdg}$& $1.1$ TeV& $1.6$ TeV& $1.6$ TeV && $0.5$ TeV\\
$Q_{cug}$& $0.9$ TeV& $1.4$ TeV& $1.3$ TeV && $0.4$ TeV\\
$Q'_{cug}$& $0.2$ TeV& $0.3$ TeV& $0.2$ TeV && $0.2$ TeV\\
\hline
\end{tabular}
\caption{Lower bounds on the average fermion resonance mass (multiplied by a combination of parameters, as indicated in the first row) in flavour anarchic models with $\tilde Y\ne 0$ (first three columns), and crude estimates in the limit $\tilde Y=0$ (last two columns). The mass bounds get stronger for larger $Y$ and/or $\tilde Y$.}
\label{tab:anarchybounds}
\end{table}

Having fixed the parametric dependences of the Wilson coefficients up to $O(1)$ factors, we can proceed to put numerical bounds on the combination $Y\tilde Y/(m_Q m_R)$ and the corresponding quantities for the subleading contributions. These bounds are listed in table~\ref{tab:anarchybounds}. We make the following observations:
\begin{itemize}
\item The strongest bounds come from the down quark (C)EDM and constrain the fermion resonance masses to be above 4--5~TeV for $Y\sim\tilde Y\sim1$.
\item For $Y\sim\tilde Y\sim1$, there is a multitude of bounds in the ballpark of 1--2~TeV. Since these refer to operators with different phases and flavour structures, we conclude that it will be hard to avoid all of them by fortuitous cancellations, even if the bounds listed here are subject to $O(1)$ uncertainties. Consequently, if $Y\sim\tilde Y$, dipole operators alone imply that sub-TeV fermion resonances are borderline and require a $Y$ not much larger than 1\footnote{%
In the anarchic case, we are only referring to the average fermion resonance masses. Individual resonances could still accidentally be much lighter without necessarily violating flavour bounds.
}.
\item In models with $\tilde Y=0$, the bounds turn out to be quite mild and an anarchic flavour and CP structure is compatible with sub-TeV fermion resonances for $Y\lesssim3$, if only constraints from dipole operators are considered.
\end{itemize}

\subsection{Flavour-symmetric models}

Since the flavour anarchic model is not only plagued by strong constraints from dipole operators but also from meson-antimeson mixing induced at tree level, it has been suggested that the strong sector is invariant under a flavour symmetry that is only broken by the composite-elementary mixings of one chirality. 
The simplest case is a $U(3)^3$ symmetry broken by the composite-elementary mixings of right-handed quarks (``left-handed compositeness'') or of left-handed quarks (``right-handed compositeness'') \cite{Barbieri:2008zt,Cacciapaglia:2007fw,Redi:2011zi}. Among the three models considered here, right-handed compositeness can only be realized in the bidoublet model, as it requires different mixings for left-handed up- and down-type quarks. While the $U(3)^3$ models successfully suppress FCNCs, they are strongly constrained by electroweak and dijet constraints, since they predict a significant degree of compositeness for one chirality of light quarks \cite{Redi:2011zi,Barbieri:2012tu}. This problem is avoided in models with a $U(2)^3$ flavour symmetry in the strong sector, again broken only by one chirality of composite-elementary mixings \cite{Barbieri:2012uh,Barbieri:2012tu}.

\subsubsection{EDM constraints in $U(2)^3$ and $U(3)^3$ models}

In $U(3)^3$ flavour models with left- or right-handed compositeness, the parameters in the strong Lagrangian are generation invariant, e.g.\ for the triplet model,
\begin{align}
(m_Q)_{ij} &= m_Q \,\delta_{ij}
\,,&
(m_R)_{ij} &= m_R \,\delta_{ij}
\,,&
(Y)_{ij} &= Y \,\delta_{ij}
\,,&
(\tilde Y)_{ij} &= \tilde Y \,\delta_{ij}
\,,
\end{align}
and analogously for the other models. It can be shown that in all models the only physical phases apart from the CKM phase reside in the wrong-chirality Yukawa couplings $\tilde Y$ \cite{Redi:2011zi}. 
In flavour models based on a $U(2)^3$ symmetry, one has instead
\begin{equation}
m_Q = \text{diag}(m_Q,m_Q,m_{Q3}) \,, \qquad
Y = \text{diag}(Y,Y,Y_3) \,,
\end{equation}
etc. As a result, there is an additional phase in the composite-elementary mixings related to the flavour symmetry-breaking spurions, but in the strong sector it is true as well that the only physical phases can be chosen to be the ones of the $\tilde Y$ couplings, which can be different for the third and the first two generations. 
Below, we will adopt a phase convention where $Y$ is real.

Consequently, in both $U(3)^3$ and $U(2)^3$ models, there is a clear-cut prediction for the flavour-conserving first-generation Wilson coefficients relevant for (C)EDMs,
\begin{align}
C_{qqV} &= a_{qV} \frac{Y \tilde{Y}}{m_{Q} m_{R}} \qquad\text{for }q=u,d
\,,
\end{align}
where in $U(2)^3$, the masses and Yukawa couplings refer to those of the first two generations of composite fermions.
Note that, in contrast to the anarchic model above, we have used a ``$=$'' sign, since there is no further $O(1)$ factor in front.
This leads to the bounds on the combination $\frac{Y \,\text{Im}\tilde Y}{m_Qm_R}$ shown in
table~\ref{tab:boundsEDM}.
We conclude that sub-TeV fermion resonances in $U(3)^3$ models, or sub-TeV fermion resonances of the first two generations in $U(2)^3$ models, require
\begin{align}
Y \, \text{Im} \tilde Y \lesssim 0.05 \,.
\end{align}

As discussed above, in the limit $\tilde Y\to 0$ the strong sector carries no new phase both in $U(3)^3$ and $U(2)^3$ models. The remaining contributions to the EDMs involving the phases in the composite-elementary mixings are tiny, and hence there is no relevant bound.

In the case of the $U(2)^3$ model, if $\tilde Y=0$ or the first generation fermion partners are decoupled, the leading contribution to the up- and down quark (C)EDMs is absent. But also the third generation wrong-chirality Yukawa $\tilde Y_3$ can contribute to the neutron EDM, if it is complex.
On the one hand, it will generate a contribution to the top CEDM via the Wilson coefficient
\begin{align}
C_{ttg} &= a_{ug} \frac{Y_3 \tilde{Y_3}}{m_{Q3} m_{R3}}
\,,
\end{align}
which leads to the bound shown in the last row of table~\ref{tab:boundsEDM}.
On the other hand, a two-loop contribution to the {\em first-generation} EDMs proportional to $\text{Im}(\tilde Y_3)$ can arise, e.g.\ from the diagram in fig.~\ref{fig:BarrZee}. Estimating this contribution naively as $C_{qqV}\sim g^2 Y \tilde Y / (16\pi^2m_\psi^2)$, one would obtain a similar bound on $m_\psi$ of the order of $0.4$~TeV for $Y_3\sim\text{Im}(\tilde Y_3)\sim 1$.

\begin{table}[tbp]
\renewcommand{\arraystretch}{1.4}
\centering
\begin{tabular}{cccc}
\hline
operator & doublet & triplet & bidoublet \\
\hline
$Q_{ddV}$& $3.6$ TeV& $5.1$ TeV& $4.1$ TeV\\
$Q_{uuV}$& $1.3$ TeV& $0.6$ TeV& $1.4$ TeV\\
$Q_{ccg}$& $1.1$ TeV& $1.7$ TeV& $1.5$ TeV \\
$Q_{bbg}$& $0.6$ TeV& $0.9$ TeV& $0.8$ TeV \\
$Q_{ttg}$& $0.3$ TeV& $0.4$ TeV& $0.4$ TeV \\
\hline
\end{tabular}
\caption{Bounds from the neutron EDM on the quantity $\sqrt{m_Qm_R}/\sqrt{Y\,\text{Im} \tilde Y}$ in $U(3)^3$ and $U(2)^3$ models.}
\label{tab:boundsEDM}
\end{table}

\subsubsection{Flavour violation in $U(3)^3$ models}

The leading contributions to the flavour-changing dipole operators in (\ref{eq:CLO}) 
vanish in models with $U(3)^3$ flavour symmetry and left- or right-handed compositeness. Subleading contributions to the unprimed Wilson coefficients arise, as discussed in section~\ref{sec:subleading}. The strongest bound is on the coefficient $C_{bsV}$, 
for which a crude estimate yields
\begin{align}
\delta C_{bsV} &\sim V_{ts} \frac{Y^2}{m_Q^2} \frac{s_{Rt}}{s_{Lt} Y} 
\,,
\end{align}
leading to the bound
\begin{equation}
 \frac{Y^2}{m_Q^2} \frac{s_{Rt}}{s_{Lt} Y} 
  \lesssim \left(\frac{1}{0.6\,\text{TeV}}\right)^2.
\label{eq:boundu3}
\end{equation}

\subsubsection{Flavour violation in $U(2)^3$ models}

In $U(2)^3$ flavour models with left-handed compositeness, the leading contributions to the Wilson coefficients read
\begin{align}
C_{bqV} &= V_{tb} V_{tq}^* \left[
a_{dV}
\left(\frac{Y \tilde{Y}}{m_{Q} m_{R}}
-\frac{Y_3 \tilde{Y}_3}{m_{Q3} m_{R3}}\right)
\right]
\qquad\text{for }q=d,s
\,,
\end{align}
and all other coefficients are negligible. Again, there are no additional $O(1)$ factors.
Since the coefficients relevant for $b\to d$ and $b\to s$ transitions are correlated in these models, it is sufficient to quote the (stronger) bound derived from the $B\to X_s\gamma$ branching ratio. It is shown in table~\ref{tab:boundsU2LC}.

In $U(2)^3$ models with right-handed compositeness, the Wilson coefficients vanish at leading order in the composite-elementary mixings. Beyond the leading order, there are contributions both in left- and right-handed compositeness analogous to the ones in $U(3)^3$ models. They give rise to a bound similar to eq.~(\ref{eq:boundu3}).

\begin{table}[tbp]
\renewcommand{\arraystretch}{1.4}
\centering
\begin{tabular}{cccc}
\hline
operator & doublet & triplet & bidoublet \\
\hline
$Q_{bsV}$& $0.37$ TeV& $0.52$ TeV& $0.22$ TeV\\
\hline
\end{tabular}
\caption{Bound on the quantity $\left(\frac{Y \tilde{Y}}{m_{Q} m_{R}}
-\frac{Y_3 \tilde{Y}_3}{m_{Q3} m_{R3}}\right)^{-1/2}$ in $U(2)^3$ flavour models with left-handed compositeness.}
\label{tab:boundsU2LC}
\end{table}

\section{Comparison with the literature}
\label{sec:comp}

Since some of the dipole operators have been considered in the literature in various models similar to the ones we studied here, we present below a detailed comparison of our findings with those of existing analyses. We find mostly agreement, but also some important differences.
\begin{itemize}
\item 
In ref.~\cite{Agashe:2008uz}, the $B\to X_s\gamma$ branching ratio has been calculated in a model similar to our doublet model, the difference being that the right-handed quarks do not mix with an $SU(2)_R$ doublet, but with two singlets, such that the strong Yukawa couplings explicitly break custodial symmetry. For the leading-order contribution to the Wilson coefficient from $W$, $Z$ or Higgs loops, this difference is however irrelevant. Up to an overall sign, we agree with the result for the charged Goldstone ($W$) contribution ($(a_{d\gamma})_W=5/12$ in our language), but disagree with the result for the neutral contribution (we find $(a_{d\gamma})_h=-1/8$ and $(a_{d\gamma})_Z=-1/24$).
\item 
In ref.~\cite{Gedalia:2009ws}, the $B\to X_s\gamma$ branching ratio and the observable $\epsilon'/\epsilon$ have been estimated in a Randall-Sundrum framework. In the anarchic doublet model, which most closely resembles their setup, the bounds we obtain from these processes are consistent with the ones found in that reference.
\item 
Ref.~\cite{Redi:2011zi} has given an estimate of the leading contribution to the $B\to X_s\gamma$ branching ratio from loops with a Higgs boson or a charged or neutral Goldstone boson ($W$ or $Z$), corresponding to our eq.~(\ref{eq:CLO}), in the anarchic bidoublet model.
We disagree with the fact that the leading-order contribution does not involve the wrong-chirality Yukawa couplings. In the same reference, EDMs in $U(3)^3$ models with left-handed compositeness were discussed, and it was claimed that the new CP-violating phase does not enter the EDM at leading order, since it can be shifted to $\tilde Y$. Our analysis shows that the converse is true: the leading contribution to the EDM is proportional to $\tilde Y$. The bounds we obtain are shown in table~\ref{tab:boundsEDM}.
\item 
The authors of ref.~\cite{Blanke:2012tv} have performed a 5D calculation of $b\to q\gamma$ processes in a Randall-Sundrum setup. The choice of fermion representations is similar to our triplet model, but the right-handed up-type quarks couple to a singlet. Furthermore, the model effectively has $\tilde Y=Y$. Our numerical estimates for the bound from the $B\to X_{s,d}\gamma$ branching fractions 
are compatible with the numerical analysis presented in that work.
\item 
Ref.~\cite{Vignaroli:2012si} contains a thorough analysis of the $B\to X_s\gamma$ branching ratio and the observable $\epsilon'/\epsilon$, closely following \cite{Agashe:2008uz}, in the triplet and bidoublet models (denoted TS5 and TS10, respectively) with flavour anarchy, setting $\tilde Y=Y$. While we agree on the overall form of the results, we have several differences in the coefficients $a_{qV}$. We present the details of our calculation in the bidoublet model in appendix~\ref{app:example}.
\item 
In ref.~\cite{Delaunay:2012cz}, the $c\to ug$ dipole transition was considered in Randall-Sundrum models in the context of $\Delta A_\text{CP}$ in $D\to KK,\pi\pi$ decays. In particular, the authors emphasize the dependence of the leading contribution on the wrong-chirality Yukawa coupling, and we confirm their findings in our 4D setup.
\end{itemize}

\section{Conclusions}
\label{sec:conc}

Dipole operators with quarks and an on-shell photon or gluon are generated at the one-loop level in theories based on the mechanism of partial compositeness, where the quarks get their masses by mixing with heavy vector-like ``composite'' fermions. Paradigm examples in this class of models are composite Higgs models or warped extra dimensions. The dipole operators contribute to numerous observables, like EDMs or FCNC decays, which can then be used to constrain these models. In this paper, we have performed an analysis of all dipole operators in the quark sector that are constrained by experiment within a simple four-dimensional setup with a single set of vector-like fermions. We have chosen this simple framework so as to be able to study the effects of choosing different representations for the composite fermion fields and of imposing different flavour structures in the strong sector. Our main findings can be summarized as follows:
\begin{itemize}
\item 
The leading contributions to the Wilson coefficients, discussed in section~\ref{sec:leading}, typically come from diagrams with a heavy fermion and a $W$, $Z$ or Higgs in the loop. These contributions are proportional to the ``wrong-chirality'' Yukawa couplings $\tilde Y$.
\item 
Beyond these leading contributions there exist a number of subleading effects, which we have categorized systematically and discussed qualitatively in section~\ref{sec:subleading}. They can be relevant, e.g., in models where the wrong-chirality Yukawas are absent or suppressed. In the case of $b\to s,d$ transitions, these subleading contributions can be comparable to the leading ones (due to the sizable degrees of compositeness of the top quark), while in all other cases they are typically suppressed by at least an order of magnitude for TeV-scale resonances.
\item 
In models with anarchic flavour and CP structures, the neutron EDM leads to a stringent constraint. If the average Yukawa couplings $Y$ and wrong-chirality Yukawa couplings $\tilde Y$ are of $O(1)$, this implies that the average fermion resonance mass scale should be above 4~TeV or so. For larger Yukawa couplings, the bounds become even stronger. Apart from the neutron EDM bound, there is a multitude of bounds in the 1--2~TeV ballpark, as summarized in table~\ref{tab:anarchybounds}.
\item 
In models in which the wrong-chirality Yukawa couplings are absent, the bounds from dipole operators are mild, even for an anarchic flavour and CP structures.
\item 
In models featuring a $U(3)^3$ flavour symmetry broken only by left- or right-handed compo\-site-elementary mixings, there is a bound from the neutron EDM that is as strong as in the anarchic case. It can be avoided by assuming the wrong-chirality Yukawa coupling to be real (or absent).
\item 
In models with a $U(2)^3$ flavour symmetry broken only by left- or right-handed composite-elementary mixings, the EDM bound can be avoided alternatively by raising the mass of the composite fermions of the first two generations.
\item 
In $U(2)^3$ flavour models with left-handed compositeness, there are bounds from flavour-violating dipoles, which are however very mild.
\end{itemize}

While our results have been obtained in the simple framework defined in section~\ref{sec:setup}, in more complete models the details of the analysis can be different. Here we only briefly comment on models where the Higgs is a pseudo Nambu-Goldstone boson (PNGB) \cite{Contino:2003ve,Agashe:2004rs}, which are particularly well motivated in view of the lightness of the Higgs boson.
Strictly speaking, these models are not a special case of the Lagrangian defined in section~\ref{sec:setup}. Given that the dominant contributions to the dipole operators come from diagrams with a heavy fermion and a Higgs, $W$ or $Z$ and does not depend on the details of the composite spin-1 sector, it is instructive to compare the fermion mass matrices in the two cases. For example, in the minimal composite Higgs model referred to as MCHM5 \cite{Contino:2006qr}, the fermion mass matrix can be written in a form (see e.g.\ \cite{Delaunay:2013iia}) that, to leading order in the expansion in the Higgs vacuum expectation value, corresponds to the mass matrix in our bidoublet model (see appendix~\ref{app:massMatrices}) with $\tilde Y=Y$.
Thus we expect that with the identification $\tilde Y=Y$ our results for the leading contributions to the Wilson coefficients also hold in composite PNGB models, up to $O(1)$ factors.

There are several ways to extend our analysis. Also in the charged lepton sector, dipole operators arise and contribute to  $\ell_i\to\ell_j\gamma$ decays, to the electron EDM or to the muon anomalous magnetic moment \cite{Agashe:2006iy,Agashe:2009tu,Redi:2013pga} (see also \cite{Kannike:2011ng,Dermisek:2013gta,Falkowski:2013jya}). 
Finally, a global numerical analysis of all contributions to $\Delta F=1$ and $\Delta F=2$ processes, taking into account electroweak constraints, would be interesting. We leave this to a future publication.

\section*{Acknowledgments}

We thank Dario Buttazzo, Michele Redi, Filippo Sala, Andrea Tesi, and Natascia Vignaroli for useful discussions. This research has been supported by the Advanced Grant EFT4LHC of the European Research Council (ERC), the Cluster of Excellence {\em Precision Physics, Fundamental Interactions and Structure of Matter\/} (PRISMA -- EXC 1098), and grant 05H12UME of the German Federal Ministry for Education and Research (BMBF). The research of D.S.\ is supported by the DFG cluster of excellence ``Origin and Structure of the Universe''.

\appendix

\section{Fermion mass matrices}
\label{app:massMatrices}

In this section we list the mass matrices of the heavy fermion resonances in all three models. In the doublet model, they are given by
\begin{align}
M^u_{\psi}=\bordermatrix{
 & t_R & T_R & U_R \cr 
t_L & 0 & -\lambda_L & 0 \cr 
T_L & 0 & m_Q & -\frac{Y v}{\sqrt{2}} \cr
U_L & -\lambda_{Ru} & -\frac{\tilde{Y} v}{\sqrt{2}} & m_R} ~,\qquad 
M^d_\psi = \bordermatrix{
 & b_R & B_R & D_R \cr
b_L & 0 & -\lambda_L & 0 \cr 
B_L & 0 & m_Q & -\frac{Y v}{\sqrt{2}} \cr 
D_L & -\lambda_{Rd} & -\frac{\tilde{Y} v}{\sqrt{2}} & m_R }~.
\end{align}
In the triplet model, they are
\begin{align}
M^u_\psi = \bordermatrix{
 & t_R & U_R & U'_R & T_R & T_{2/3 R} \cr
t_L & 0 & 0 & 0 & -\lambda_L & 0 \cr
U_L & -\lambda_{Ru} & m_R & 0 & -\frac{\tilde{Y}v}{2} & \frac{\tilde{Y}v}{2} \cr
U'_L & 0 & 0 & m_R & -\frac{\tilde{Y}v}{2} & \frac{\tilde{Y}v}{2} \cr
T_L & 0 & -\frac{Yv}{2}& -\frac{Yv}{2} & m_Q & 0 \cr
T_{2/3 L} & 0 & \frac{Yv}{2} & \frac{Yv}{2} &0 & m_Q}\,, \qquad
M^d_{\psi} = \bordermatrix{
 & b_R & D_R & D'_R & B_R \cr 
b_L & 0 & 0 & 0 & -\lambda_L \cr
D_L & -\lambda_{Rd} & m_R & 0 & -\frac{\tilde{Y}v}{\sqrt{2}} \cr 
D'_L & 0 &0 & m_R & -\frac{\tilde{Y}v}{\sqrt{2}} \cr
B_L & 0 & -\frac{Yv}{\sqrt{2}} & -\frac{Yv}{\sqrt{2}} & m_Q}\,, 
\end{align}
and
\begin{equation}
M^{5/3}_\psi = \bordermatrix{
 & T_{5/3R} & U_{5/3R} & U'_{5/3R} \cr
T_{5/3L} & m_Q & -\frac{Y v}{\sqrt{2}} & -\frac{Yv}{\sqrt{2}}\cr
U_{5/3L} & -\frac{\tilde{Y}v}{\sqrt{2}} & m_R & 0 \cr
U'_{5/3L} & -\frac{\tilde{Y}v}{\sqrt{2}} & 0 & m_R}\,.
\end{equation}
In the bidoublet model, they are
\begin{equation}
\begin{aligned}
M^u_\psi &= \bordermatrix{
 & t_R & T_R & T'_R & T_{2/3 R} & U_R \cr
t_L & 0 & -\lambda_{Lu} & -\lambda_{Ld} & 0 & 0 \cr
T_L & 0 & m_{Qu} & 0 & 0 & -\frac{Y v}{\sqrt{2}} \cr
T'_L & 0 & 0 & m_{Qd} & 0 & 0 \cr 
T_{2/3 L} & 0 & 0 & 0 & m_{Qu} & -\frac{Y v}{\sqrt{2}} \cr
U_L & -\lambda_{Ru} & -\frac{\tilde{Y} v}{\sqrt{2}} & 0 & -\frac{\tilde{Y} v}{\sqrt{2}} & m_U} \,, \\ \\
M^d_\psi&=\bordermatrix{
 & b_R & B_R & B'_R & B_{-1/3 R} & D_R \cr
b_L & 0 & -\lambda_{Lu} & -\lambda_{Ld} & 0 & 0 \cr
B_L & 0 & m_{Qu} & 0 & 0 & 0 \cr 
B'_L & 0 & 0 & m_{Qd} & 0 & -\frac{Y v}{\sqrt{2}} \cr
B_{-1/3 L}  &  0  &  0  & 0 & m_{Qu} & -\frac{Y v}{\sqrt{2}} \cr 
D_L & -\lambda_{Rd} & 0 & -\frac{\tilde{Y} v}{\sqrt{2}} & -\frac{\tilde{Y} v}{\sqrt{2}} & m_D} \,.
\end{aligned}
\end{equation}
In the discussion of our results we have switched from the mixing parameters $\lambda_i$ to the sines of the mixing angles determining the degree of compositeness. The composite-elementary mixings are in general given by $s_i\equiv\sin\varphi_i\approx\tan\varphi_i= \lambda_i/m_i$; more specifically, for the doublet and triplet models
\begin{align}
s_{Lt}=s_{Lb} \equiv s_L = \frac{\lambda_L}{\sqrt{m_Q^2+\lambda_L^2}} \,, \quad 
s_{Rt} = \frac{\lambda_{Ru}}{\sqrt{m_R^2+\lambda_{Ru}^2}} \,, \quad 
s_{Rb} = \frac{\lambda_{Rd}}{\sqrt{m_R^2+\lambda_{Rd}^2}} \,,
\end{align}
whereas in the bidoublet model
\begin{align}
s_{Lt} = \frac{\lambda_{Lu}}{\sqrt{m_{Qu}^2+\lambda_{Lu}^2}} \,, \quad 
s_{Rt} = \frac{\lambda_{Ru}}{\sqrt{m_{U}^2+\lambda_{Ru}^2}} \,, \quad
s_{Lb} = \frac{\lambda_{Ld}}{\sqrt{m_{Qd}^2+\lambda_{Ld}^2}} \,, \quad 
s_{Rb} = \frac{\lambda_{Rd}}{\sqrt{m_{D}^2+\lambda_{Rd}^2}} \,.
\end{align}

\section{Model-independent formulae for the Wilson coefficients}
\label{app:reference}

\begin{table}
\renewcommand{\arraystretch}{1.4}
\centering
\begin{tabular}{cccc||cccc}
\hline 
$\mathcal{C}$ & $X$  & $Q_F$ & $Q_X$& $\mathcal{C}$ & $X$  & $Q_F$ & $Q_X$ \\ 
\hline 
$\mathcal C_{dd\gamma}$ & $h, Z, \rho^0$ & $-1/3$ & $0$ & $\mathcal C_{uu\gamma}$& $h,Z,\rho^0$ & $2/3$ & $0$\\ 
  & $W^-, \rho^-$ & $-4/3$ & $-1$ &  & $W^-, \rho^-$ & $-1/3$ & $-1$\\ 
  & $W^+, \rho^+$ & $2/3$ & $1$ & & $W^+, \rho^+$ & $5/3$ & $1$\\ 
  & $G^\ast$ & $-4/9$ & $0$ &  & $G^\ast$ & $8/9$ & $0$\\ 
\hline 
$\mathcal C_{ddg}$ & $h, W, \rho$ & $1$ & $0$ & $\mathcal C_{uug}$& $h,W,\rho$ & $1$ & $0$\\ 
  & $G^\ast$ & $-1/6$ & $3/2$ &  & $G^\ast$ & $-1/6$ & $3/2$\\ 
\hline 
\end{tabular} 
\caption{Charge parameters for the loop functions (\ref{eq:LoFus}) depending on contribution}\label{tab:LoFuArguments}
\end{table}

Here we give the exact analytical expressions for the one-loop Wilson coefficients, which were used to obtain the approximate expressions given in the text. The Wilson coefficients of the $q_i\to q_j\gamma$ dipole operators, as defined in eqs.~(\ref{eq:Heff})--(\ref{eq:O2}) with $i>j$, can be written as
\begin{align}
\mathcal{C}_{q_i q_j \gamma,g} &= \sum \limits_{\psi,\, X} \frac{1}{m_{q_i} m_X^2}\left(m_{q_i} V^{L\ast}_{i\psi X} V^L_{j\psi X} + m_{q_j} V^{R\ast}_{i\psi X} V^R_{j\psi X}\right) F^1_X (Q_\psi, Q_X, x) \\ 
&\quad\mbox{}+ \frac{1}{m_{q_i} m_X^2} \left(m_\psi V^{L\ast}_{i\psi X} V^R_{j\psi X}\right) F^2_X (Q_\psi,Q_X,x)\,, \label{eq:AppWilson}
\end{align}
where $\psi$ denotes the fermion entering the loop and $X$ can be either vector, scalar or a heavy gluon resonance. The parameter $x$ is given by $x = m_\psi^2/m_X^2$.
The expression for the primed Wilson coefficient $\mathcal{C}_{q_i q_j \gamma,g}'$ can be obtained from (\ref{eq:AppWilson}) by interchanging $L\leftrightarrow R$.
The loop functions are defined as
\begin{align}
F_V^1(Q_\psi,Q_V,x) &=
Q_\psi\frac{\left(5 x^4-14 x^3+39 x^2-18 x^2 \log x-38 x+8\right)}{24 (x-1)^4} \nonumber\\
&+Q_V\frac{\left(4 x^4-49 x^3+18 x^3 \log x + 78 x^2-43 x+10\right)}{24 (x-1)^4}\,,  \\
F_V^2(Q_\psi,Q_V,x) &= 
Q_\psi\frac{ \left(-x^3-3 x+6 x \log x+4\right)}{4 (x-1)^3} 
 +Q_V\frac{ \left(-x^3+12 x^2-6 x^2 \log x-15 x+4\right)}{4 (x-1)^3}\,, \\
F_S^1(Q_\psi,Q_S,x) &=
Q_\psi \frac{\left(-x^3+6 x^2-3 x-6 x \log x-2\right)}{24 (x-1)^4}\nonumber\\
&+Q_S\frac{ \left(2 x^3+3 x^2-6 x^2 \log x-6 x+1\right)}{24 (x-1)^4} \,, \\
F_S^2(Q_\psi,Q_S,x) &= Q_\psi \frac{\left(-x^2+4 x-2 \log x-3\right)}{4 (x-1)^3} +Q_S \frac{\left(x^2-2 x \log x-1\right)}{4 (x-1)^3} \,.
\label{eq:LoFus}
\end{align}
Note that the charge parameters $Q_{\psi,V,S}$ in the loop functions are not necessarily the electric charges of the corresponding particles but can be color factors in the cases of either the external gauge field being a gluon or the loop involving heavy gluon resonances. A complete reference is given in table~\ref{tab:LoFuArguments}. The couplings $V_{i\psi X}^{L/R}$ in eq.~(\ref{eq:AppWilson}) are defined as follows:
\begin{align}
\Diagram{\vertexlabel^{q_i} \\
fdV \\
& g \vertexlabel_{V} \vertexlabel^{\mu}\\
 fuA\\
 \vertexlabel_{\psi}
}~ &=\, i \bar q_i \gamma^\mu\left(V_{i\psi V}^L P_L + V_{i\psi V}^R P_R \right) V_\mu \psi  \nonumber \\
 \nonumber \\
\Diagram{\vertexlabel^{q_i} \\
fdV \\
& h \vertexlabel_{S}\\
 fuA\\
 \vertexlabel_{\psi}
} ~ &= \, i \bar q_i \left(V_{i\psi S}^L P_L + V_{i\psi S}^R P_R\right) S \psi  \nonumber \\
 \nonumber \\
\Diagram{\vertexlabel^{q_i,\, \alpha} \\
fdV \\
& g \vertexlabel_{G^\ast} \vertexlabel^{a,\mu}\\
 fuA\\
 \vertexlabel_{\psi,\, \beta}
}~ &=\, i \bar q_{i}^\alpha \left( V_{i \psi G}^L P_L + V_{i \psi G}^R P_R\right) {G^\ast_\mu}^a \psi^\beta T^a_{\alpha \beta}
\end{align}

\section{Calculation of leading contribution in the bidoublet model}
\label{app:example}

\begin{figure}
\centering
\begin{subfigure}[b]{0.24\textwidth}
\includegraphics[scale=0.65]{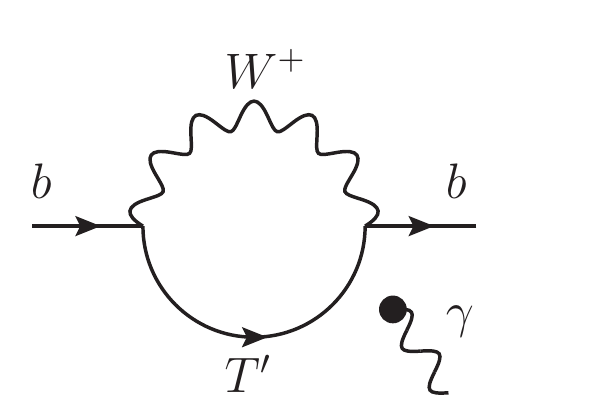} 
\end{subfigure}
\begin{subfigure}[b]{0.24\textwidth}
\includegraphics[scale=0.65]{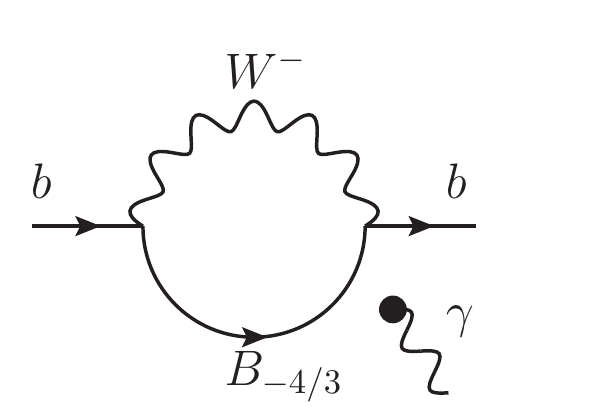} 
\end{subfigure}
\begin{subfigure}[b]{0.24\textwidth}
\includegraphics[scale=0.65]{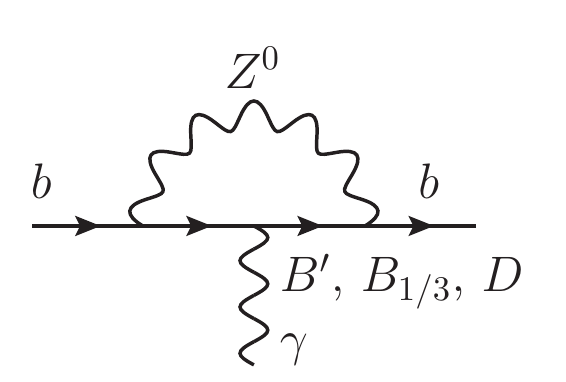} 
\end{subfigure}
\begin{subfigure}[b]{0.24\textwidth}
\includegraphics[scale=0.65]{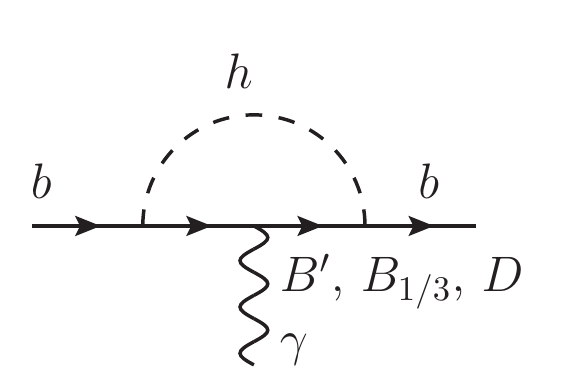} 
\end{subfigure}
\caption{\label{fig:graphsxxx}
Diagrams contributing to the leading correction in the bidoublet model. The blob on the photon leg denotes the photon either coupling to the loop fermion or the $W$ boson. We use the same names for the fermion mass eigenstates as for the fermion fields in the composite-elementary basis. The mass eigenstates are understood to correspond to the fields in the original basis with whom they have the largest admixtures.
}
\end{figure}

Here we illustrate the calculation for the leading-order correction to $\mathcal C_{qq\gamma}$ in the bidoublet model for one generation. This contribution is governed by diagrams with a heavy fermion and a $W$, $Z$ or Higgs in the loop, as shown in fig.~\ref{fig:graphsxxx}.
We obtain
\begin{align}
\mathcal C_{bb\gamma} =\sum \limits_{\psi, X} \frac{m_\psi}{m_b m_X^2} V_{b\psi X}^R V_{b \psi X}^{L*} \, F_X \left(Q_\psi, Q_X, x\right) ,
\end{align}
where $m_\psi$ and $m_X$ are the masses of the fermion and the boson in the loop. The $V_{b \psi X}^{L,R}$ are the fermion-gauge couplings in the mass eigenbasis, as defined in app.~\ref{app:reference}. For the loop functions, we use the approximations
\begin{equation}
F_V \left(Q_\psi,Q_V,x\right) \approx -\frac{Q_\psi+Q_V}{4}  \,, \quad
F_S \left(Q_\psi,x\right) \approx -\frac{Q_\psi}{x}\,,
\end{equation}
where we only kept the first non-vanishing order for $x\to\infty$. The gauge couplings up to quadratic order of the composite elementary mixings are given in table~\ref{tab:appCouplings}. Table~\ref{tab:appMasses} lists the mass eigenstates to order $O(v)$. We have followed the convention of setting $m_{Qu}=m_{Qd}\to m_Q$ and $m_U=m_D\to m_R$ everywhere. Putting all the pieces together, we find the contributions listed in table~\ref{tab:appContrib}. Summing up these contributions, we have obtained the result from section \ref{sec:WilsonAnalytic}, i.e.\
\begin{align}
\mathcal C_{bb\gamma} = \sum \limits_i c_i = a_{d\gamma}\frac{Y\tilde Y}{m_Q m_R}\,,
\end{align}
with $a_{d\gamma} = -1/2$.

\begin{table}
\renewcommand{\arraystretch}{2}
\centering
\begin{tabular}{cccc}
\hline 
$\psi$& $X$ & $V_{b\psi X}^R$ & $V_{b\psi X}^L$ \\ 
\hline
$T^\prime$ & $W^-_\mu$ & $\dfrac{g Y \tilde Y v^2 s_{Lb}}{2 \sqrt{2}m_R m_Q}$ & $-\dfrac{g Y v s_{Rb}}{2 m_Q} $ \\
\hline
$B_{-4/3}$ & $W^+_\mu$ & $\dfrac{g Y \tilde Y v^2 s_{Lb}}{2\sqrt{2} m_Q m_R} $ & $-\dfrac{g Y v s_{Rb}}{2 m_Q} $ \\
\hline
$B^\prime$ & $Z^0_\mu$ & $\dfrac{g Y v s_{Rb}}{2\sqrt{2} m_Q \sqrt{1-s_W^2}}$ & $\dfrac{g Y\left(Y m_R+\tilde Y m_Q\right) v^2 s_{Lb} }{4 m_R \left(m_R^2-m_Q^2\right) \sqrt{1-s_W^2}}$ \\ 
\hline 
$B_{1/3}$ & $Z^0_\mu$ & $-\dfrac{g Y v s_{Rb}}{2\sqrt{2}m_Q \sqrt{1-s_W^2}}$ & $\dfrac{g Y \left(Y m_Q m_R -\tilde Y m_Q^2+2 \tilde Y m_R^2\right) v^2 s_{Lb}}{4 m_Q m_R \left(m_R^2-m_Q^2\right) \sqrt{1-s_W^2}}$ \\ 
\hline 
$D$ & $Z^0_\mu$ & $0$ & $-\dfrac{g Y v s_{Lb}}{2\sqrt{2}m_R \sqrt{1-s_W^2}}$ \\ 
\hline 
$B^\prime$& $h$ & $ \dfrac{Y s_{Rb}}{\sqrt{2}}$&$\dfrac{Yv \left(\tilde Y m_Q^2 - 2 \tilde Y m_R^2-Y m_Q m_R\right) s_{Lb}}{2 m_R\left(m_R^2-m_Q^2\right)}$ \\ 
\hline 
$B_{1/3}$& $h$ & $ \dfrac{Y s_{Rb}}{\sqrt{2}}$ & $\dfrac{Yv \left(\tilde Y m_Q^2 - 2 \tilde Y m_R^2-Y m_Q m_R\right) s_{Lb}}{2 m_R\left(m_R^2-m_Q^2\right)}$\\ 
\hline 
$D$&$h$&$\dfrac{Y \tilde Y v \left(m_R^2-2m_Q^2\right)s_{Rb}}{mL\left(m_Q^2-m_R^2\right)}-\dfrac{Y^2 v s_{Rb}}{m_Q^2-m_R^2}$ & $ \dfrac{Y s_{Lb}}{\sqrt{2}}$ \\ 
\hline
\end{tabular}
\caption{Couplings of the $b$ and the loop fermion to the loop boson in the bidoublet model}\label{tab:appCouplings}
\end{table}
\begin{table}
\renewcommand{\arraystretch}{2}
\centering
\begin{tabular}{cccccc}
\hline 
\rule[-1ex]{0pt}{2.5ex} $b$ & $B'$ & $B_{1/3}$ & $D$ & $T'$ & $B_{-4/3}$ \\ 
\hline 
\rule[-1ex]{0pt}{2.5ex} $\dfrac{Y v}{\sqrt{2}} s_{Lb}s_{Rb}$ & $m_Q$ & $m_Q$ & $m_R+\dfrac{m_R}{2}s_{Rb}^2$ & $m_Q$ & $m_Q$ \\ 
\hline 
\end{tabular} 
\caption{Mass eigenstates of the involved fermions at order $O(v)$ in the bidoublet model}\label{tab:appMasses}
\end{table}

\begin{table}
\renewcommand{\arraystretch}{2.2}
\centering
\begin{tabular}{ccc}
\hline 
\rule[-1ex]{0pt}{2.5ex} Loop Fermion & Loop Boson & $c_i$ \\ 
\hline 
\rule[-1ex]{0pt}{2.5ex} $T'$ & $W^-_\mu$ & $\dfrac{5 Y \tilde Y}{12 m_Q m_R}$ \\ 
\hline 
\rule[-1ex]{0pt}{2.5ex} $B_{-4/3}$ & $W^+_\mu$ & $-\dfrac{7 Y \tilde Y}{12 m_Q m_R}$ \\ 
\hline 
\rule[-1ex]{0pt}{2.5ex} $B'$ & $Z^0_\mu$ & $\dfrac{Y \tilde Y m_Q}{24\left(m_R^3-m_Q^2 m_R\right)}$ \\ 
\hline 
\rule[-1ex]{0pt}{2.5ex} $B_{1/3}$ & $Z^0_\mu$ & $\dfrac{Y\tilde Y \left(m_Q^2 -2m_R^2\right)}{24\left(m_Q m_R^3-m_Q^3 m_R\right)}$ \\ 
\hline 
\rule[-1ex]{0pt}{2.5ex} $D$ & $Z^0_\mu$ & $0$ \\ 
\hline 
\rule[-1ex]{0pt}{2.5ex} $B'$ & Higgs & $\dfrac{Y \tilde Y \left(m_Q^2-2m_R^2\right)}{24\left(m_Qm_R^3-m_Q^3m_R\right)}$ \\ 
\hline 
\rule[-1ex]{0pt}{2.5ex} $B_{1/3}$ & Higgs & $\dfrac{Y \tilde Y \left(m_Q^2-2m_R^2\right)}{24\left(m_Qm_R^3-m_Q^3m_R\right)}$ \\ 
\hline 
\rule[-1ex]{0pt}{2.5ex} $D$ & Higgs & $\dfrac{Y\tilde Y\left(2m_Q^2-m_R^2\right)}{12\left(m_Q^3 m_R-m_Qm_R^3\right)}$ \\ 
\hline 
\end{tabular} 
\caption{Contributions relevant to the leading order correction of $C_{bb\gamma}$ in the bidoublet model}\label{tab:appContrib}
\end{table}

\newpage
\bibliographystyle{JHEP}
\bibliography{dipoles}

\end{document}